# SHORT TIME DIE ATTACH CHARACTERIZATION OF SEMICONDUCTOR DEVICES

*Péter Szabó[1,2], Márta Rencz[1,2]*

[1]Budapest University of Technology, Department of Electron Devices, Budapest XI.,
Goldman György tér 3. H-1111 Hungary, tel.:(+36-1) 463-2727

[2]MicReD Ltd, Budapest XI., Etele út 59-61, H-1119 Hungary, tel.: (+36-1) 481-1369

**ABSTRACT**

*Thermal qualification of the die attach of semiconductor devices is a very important element in the device characterization as the temperature of the chip is strongly affected by the quality of the die attach. Voids or delaminations in this layer may cause higher temperature elevation and thus damage or shorter lifetime. Thermal test of each device in the manufacturing process would be the best solution for eliminating the devices with wrong die attach layer.*

*In this paper we will present the short time thermal transient measurement method and the structure function evaluation through simulations and measurements for die attach characterization. We will also present a method for eliminating the very time consuming calibration process. Using the proposed methods even the in-line testing of LEDs can be accomplished.*

## 1. INTRODUCTION

In case of power switches, LED and stacked die devices the thermal characterisation of the die attach is inevitable to diminish the chance of failure [1][2][3]. Testing each device during the manufacturing process would be the best solution for eliminating the devices with Poor die attach layer [5].

Thermal characterization of packaged devices can be easily done by the Structure Function evaluation method [4]. To create the structure function of a device we need the thermal transient heating or cooling curve of the sample. Reaching the needed steady states at the beginning and the end of the measurement may take minutes or even longer time. The transient measurement also demands sensitivity calibration process for temperature measurement. The time interval of this process is in the range of minutes too. Unfortunately in most of the cases the method is too long to be useful for in-line (manufacturing) testing.

Measuring the thermal transient for shorter time may be usable for establishing the thermal property of the die attach as the early section of the transient describes the early section of the measured structure. The loss of information about the further parts of structure is inevitable in this type of measurement. It was experienced in our simulations and measurements that the early section of the thermal transient is really affected by the thermal property of the die attach. It was also experienced that the end of the transient is mainly affected by the thermal boundary conditions.

We also found that the structure function evaluation method can give the detailed model of the early section of the measured structure even though the transient was not measured between steady states (it was short) [5].

The only problem we found a bit astonishing in this type of measurement is the calibration procedure. In case of in line testing the exact value of the thermal resistance is not demanded. It is enough to establish the difference between the measured sample and a so called "golden sample". We could solve this problem using the similarity at the very early section of the transients between different samples from the same type. With this method we found that the needed measurement time of a device is in some 10ms time range.

## 2. THERMAL SIMULATIONS

To check the feasibility of the short time die attach characterization we simulated a silicon power device with the cross sectional view and dimensions of Figure 1. The SUNRED thermal simulator was used in this process [6]. Two simulations have been evaluated for the temperature distribution by changing the thermal conductivity of the die attach, see Figure 2 and Figure 3. In case of Figure 2 the die attach layer has low thermal conductivity representing good thermal die attach quality. In case of Figure 3 the layer has poor conductivity property modelling the presence of voids and/or delaminations, wrong thermal die attach quality.

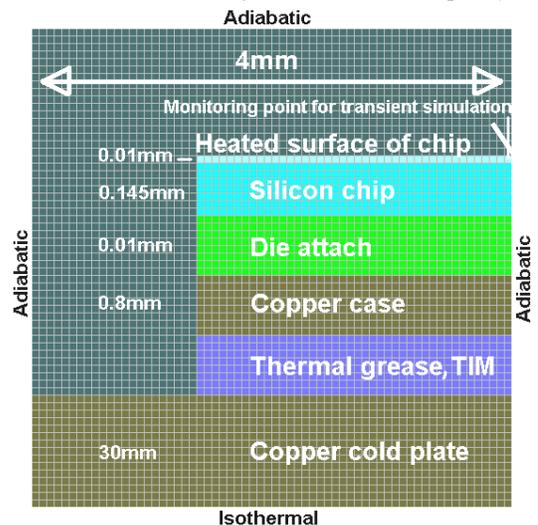

**Figure 1.** Cross sectional view of the mirror-symmetrical model, equidistant mesh

In the simulation process we began heating the structure by 1W (step-like heating) and captured the temperature distribution. We compared the temperature distribution at 1.2 ms. The simulation in Figure 3 shows larger temperature elevation at the top of the chip (heated surface). The larger die attach thermal resistance obstacles the heat.

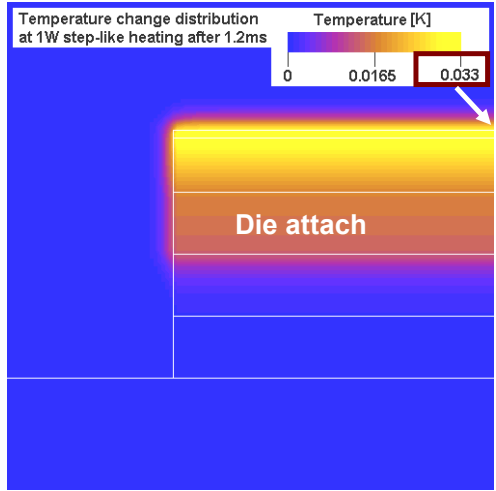

**Figure 2.** Cross sectional view of the distribution of the temperature change of the Good die attach model after 1.2ms at 1W step-like heating, equidistant mesh

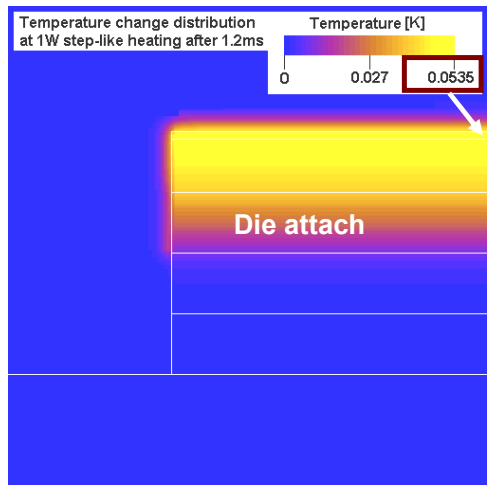

**Figure 3.** Cross sectional view of the distribution of the temperature change of the Poor die attach model after 1.2ms at 1W step-like heating, equidistant mesh, larger temperature elevation

In Figure 4 we compared the thermal impedance curves ($Z_{th}$) of the monitoring point located in the middle of the heated area of the chip (Figure 1). The curves are running together till 100µs describing the similar heat spreading in the two cases. The chip regions are the same in both cases. The curve of Poor die attach shows larger temperature elevation. The curves are running parallel after approximately 10ms as the heat spreads again in similar structures: copper case, TIM, copper cold plate.

Figure 5 shows the structure functions calculated from the thermal impedance curves. We can notice that the effect is similar to the $Z_{th}$ curves at the beginning: the heat spreading ways are similar up to approximately 0.1K/W. After this section there is a slope up to about 0.3K/W in case of the good, and up to about 0.07K/W in case of the Poor die attach. These sections show the heat spreading in the die attach layers of the two models. The Poor die attach case shows a longer horizontal section meaning larger thermal resistance. After this section the structure functions practically run parallel similarly to the $Z_{th}$ curves and meaning the heat path is similar.

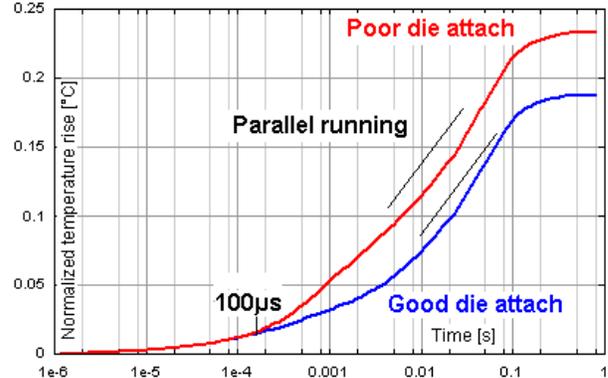

**Figure 4.** Thermal impedance curves of the Good and Poor die attach cases

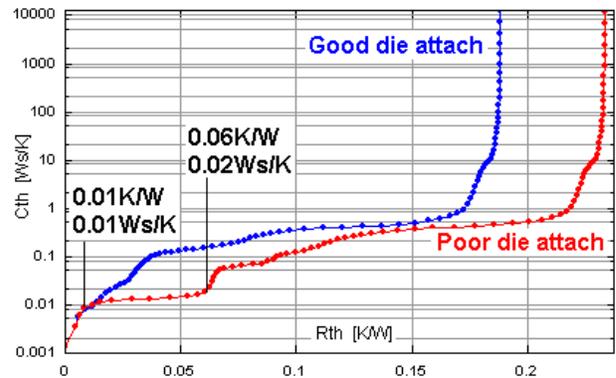

**Figure 5.** Enlarged view of the structure functions calculated from the $Z_{th}$ curves of Figure 4

For further discussions one important property of the structure function has to be taken into consideration: the characteristic time constant of any section can be easily established from the structure function [4]. Eg.: if the thermal resistance of the end of the chip section is multiplied by its thermal capacitance, then the result is the average time constant of the chip region.

The calculation for the simulation gives about 100µs:

(1)    $\tau_{chip} = R_{th} * C_{th} = 0.01K/W * 0.01Ws/K = 100µs$

This value is equals to the time value of the beginning of the separation in the $Z_{th}$ curves, see Figure 4.

The time needed for the characterisation of the die attach is in the range of the time constant of the chip+die attach section:

(2) $\quad \tau_{chip+die\ attach} = 0.06K/W * 0.02Ws/K = 1.2ms$

In Figure 4 the difference between the $Z_{th}$ curves can be seen well at 1.2ms. The difference shows the different thermal die attach quality.

### 3. MEASUREMENTS OF TO-LIKE PAKAGES

We presented above that both the Zth curve and the Cumulative structure function are applicable for the examination of the die attach quality. Let us find out how much time is needed for the short time measurement of the die attach qualities of usual packages of power devices.

Three TO type packages were measured using the steady to steady state thermal transient measurement method. We calculated the structure functions and the necessary time for short time die attach characterization. The three results are shown in Figure 6, Figure 7 and Figure 8. Two functions can be seen in each Figure. One has "No ceramic" notation, this was calculated from the transient measured in a thermal setup in which the device was pressed against the cold plate by its copper case and thermal grease was applied between the surfaces. In the other case the name is "Ceramic". In this case ceramic sheet was placed between the surfaces. As consequence of the different thermal setup the functions are also different. The point where they begin diverging corresponds to the case [7].

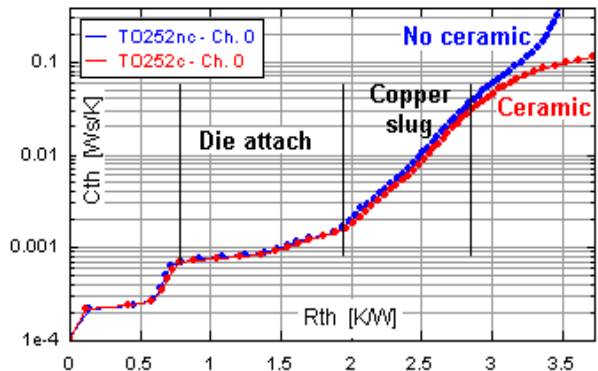

**Figure 6.** TO252 package, $\tau \approx$ 1mWs/K* 2K/W = 2ms

On the left hand side of this point a steep curving section can be found representing the heat spreading in the copper case of the package. The section before this having smaller steepness represents the die attach region. The point where the steepness is changing is the end of the die attach. Its time constant gives the minimal necessary time for the short time die attach measurement

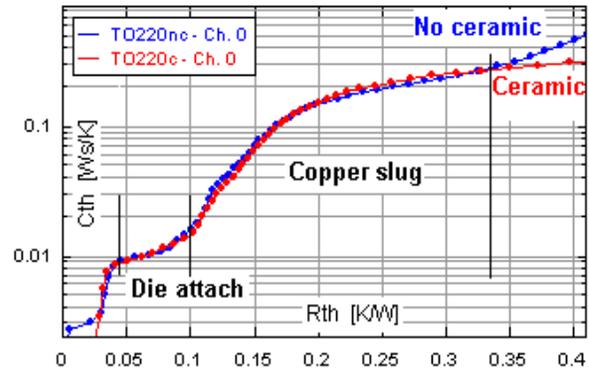

**Figure 7.** TO220, $\tau \approx$ 10mWs/K * 0.1K/W = 1ms

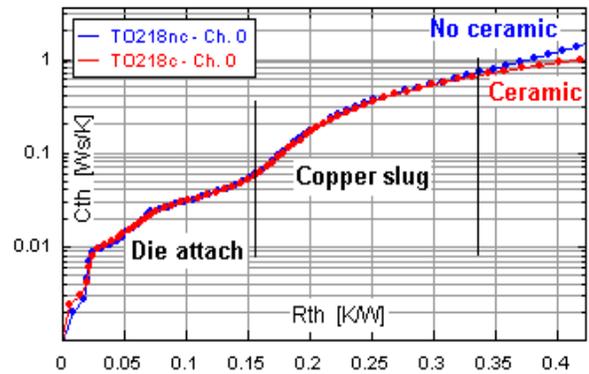

**Figure 8.** TO218, $\tau \approx$ 50mWs/K* 0.15K/W = 7.5ms

The results showed that the characteristic time of the die attach is in the ms range in case of these three typical packages. 10 ms long thermal transient measurement seems to be enough to establish the die attach quality.

### 4. ELIMINATION OF THE CALIBRATION PROCESS

The steady to steady thermal transient measurement demands several appliances: Thermal transient tester, thermostat, control computer, evaluation software [4][7]. The method is more or less the same for the short time die attach characterization, so the needed appliances are also the same: we heat up the device electronically for some ms and capture the change of the temperature dependent electrical parameter [5].

In short time die attach characterization there is an additional problem coming from the unknown sensitivity of the temperature dependent electrical parameter that has to be known for temperature calculation. The temperature sensitive parameter can be e.g.: the threshold voltage of diode or the gate-source threshold voltage of MOS device. The electron devices show different sensitivities; even they were made by the same fabrication process. For this reason in the short time die attach characterisation devices may show different voltage change not only because of different die attach quality but also because of the presence of different sensitivity values. This effect can be eliminated by using the similarity of the chip region of the devices of the same type. The thermal properties of the chips are usually very stabile as the thickness and the heat conductivity and capacity values of the single crystal silicon are very similar. So, the measured thermal transient curves have to be the same at their beginnings. If the temperature transient caused voltage transient response is different at the early region we can take the assumption that the difference is coming from the sensitivity difference. Applying a linear transformation at each measured voltage curve we can fit them at their beginnings to one curve measured at a sample having good die attach quality. After the fitting the comparison between the fitted curves can be done by the structure function evaluation method.

## 5. SHORT TIME DIE ATTACH CHARACTERIZATION OF LED SAMPLES

The power dissipated in LEDs is growing strongly nowadays. Thermal characterisation of each device would be the best even the large number of the devices in the manufacturing process [5].

Two LEDs from the same type were investigated; one had die attach problem (Poor die attach), the other was known good (Good die attach) [3]. We carried out thermal transient measurements on the samples. The devices were placed on cold plate. Approximately 0.5 W power excitation was used and the emitted optical power was measured too. The $Z_{th}$ curves are running together up to about 1ms describing the similarity of the chip region, see Figure 9. After the diverging point the divergence holds up to about 4ms and after this point the curves begin running together again.

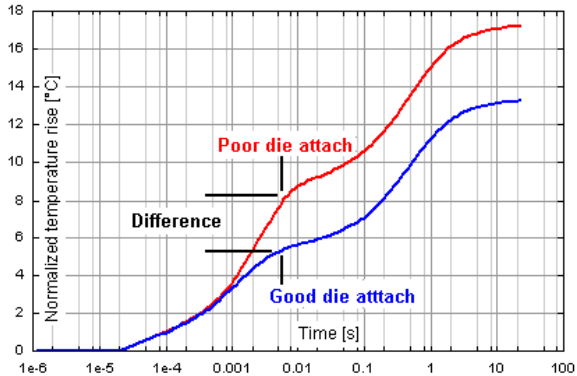

**Figure 9.** Steady to steady state $Z_{th}$ curves of the two LED samples having Good and Poor die attach layers

For better identification of the die attach regions we calculated the cumulative structure functions from the $Z_{th}$ curves. The difference presents in the die attach layer too, see Figure 10. 3K/W is the thermal resistance of the die attach of the Good device and 6K/W of the Poor device.

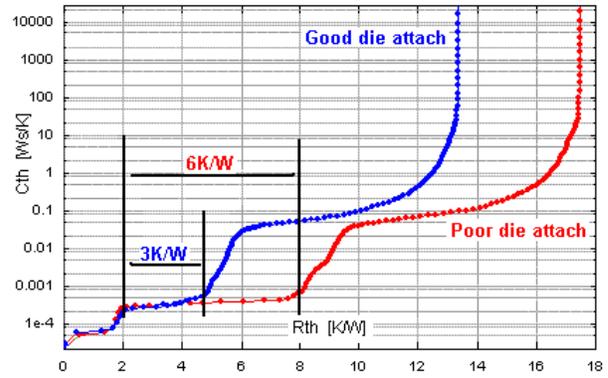

**Figure 10.** Cumulative structure functions of the two LED samples

We calculated the time needed for short time die attach characterisation of the Poor device:

(3)      $\tau_{chip+die\ attach}$ = 8K/W*0.5mWs/K =4 ms.

It correlates well with the transient results. The 4ms was the time where the curves began to run parallel.

In the next step we used a special mode of heating: the device was driven during 10 ms by the 0.5W and the thermal transient was captured during 50ms immediately after the excitation was switched off. The reason why this driving method was chosen is the smaller cross-electrical effect at the beginning of the transient after switching of the driving current. The other reasons are: the power step is more ideal after the switching off, the sensitivity calibration is easier for later comparison. This measurement is of course different from the steady to steady state one. Figure 11 shows the difference between the transients of steady to steady state and short type measurements. They are running well together up to 4ms. This type of short time driving is applicable for establishing differences in the die attach layer. At the early sections of the short $Z_{th}$ curves of the Poor and Good devices are very similar. They show the difference at 4ms well.

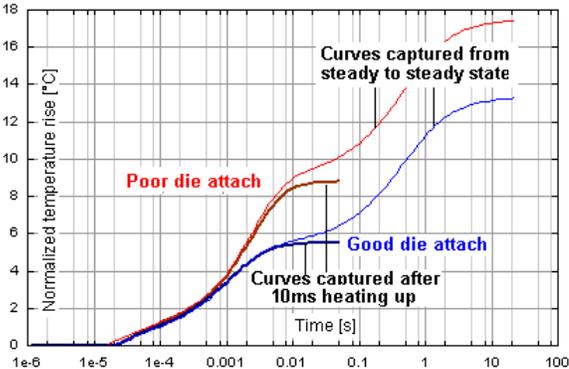

**Figure 11.** Comparison of the $Z_{th}$ curves of the steady to steady state and short time transient measurements

We calculated the structure functions from the short time transients to examine the effect of the length of the thermal transient measurement to the die attach region, see Figure 12. The structure functions calculated from the short time transient measurements are also applicable to establish the thermal resistances and capacitances of the early regions.

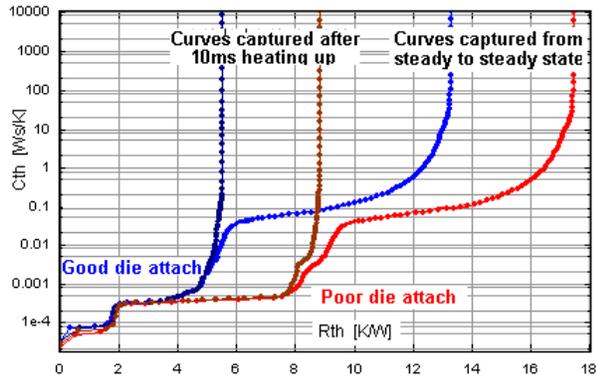

**Figure 12. Cumulative structure functions calculated from the steady to steady and short time measurements**

We have measured large number of LEDs from the same type in the same way. The temperature change transients caused threshold voltage change transients can be seen in Figure 13. The curves show large variation at their ends. The curves also show large variation at their beginnings from 10µs to 1ms that corresponds to the chip region. From these measurements we can not draw any conclusions in this form.

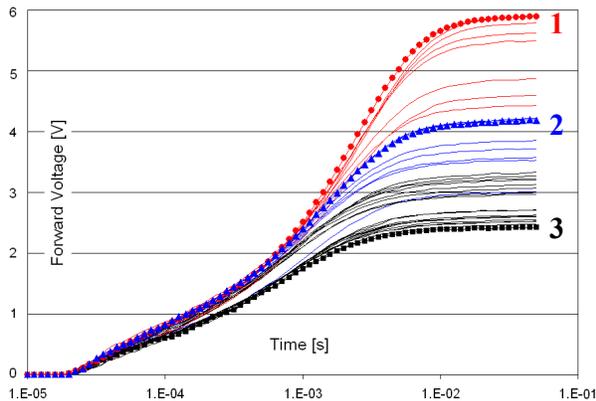

**Figure 13. Thershold voltage change of LEDs from the same type after 10ms driving**

In order to find out the origin of the variations, we measured the sensitivities of the devices and calculated the thermal impedance curves, see Figure 14. They are practically similar at their beginnings. The good fitting (similarity) at the beginnings is coming from the similarity of the chip regions of the LEDs.

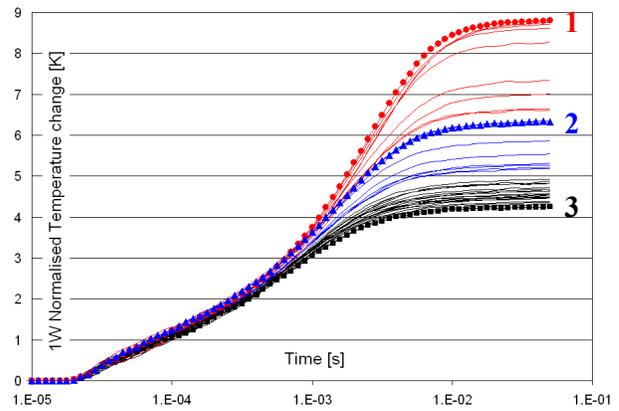

**Figure 14. $Z_{th}$ curves with good sensitivity**

These measurements proved that the difference at the beginnings is coming form the different sensitivity values.

In the next step we calculated the structure functions of the three typical curves of Figure 14 (1, 2, and 3) using same sensitivity, see Figure 15. The difference at their beginnings can be experienced well. We calculated the structure functions from the thermal transient curves using the sensitivities from the calibration process, see Figure 16. The good fit at their beginning can be seen well. We choose two points (crosses sign) in this region for linear fitting. The three structure functions of Figure 15 were fitted to these two points. The result, the fitted structure functions can be seen in Figure 17.

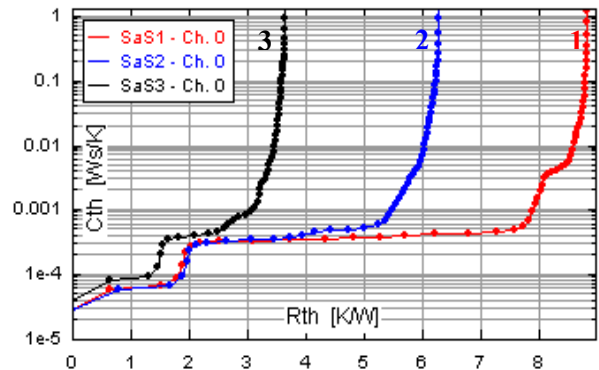

**Figure 15. Structure functions of the three characteristic samples (1,2,3) calculated from the thermal impedance curves using same sensitivity**

The fitting method gives good results as the beginning of the fitted curves are nearly the same and the order of the ends are similar to the ends of the structure functions calculated with good sensitivities. Devices having die attach problems can be dropped out by reading the thermal resistance at a given thermal capacitance, now e.g. at 1 Ws/K, and comparing to the desired thermal resistance.

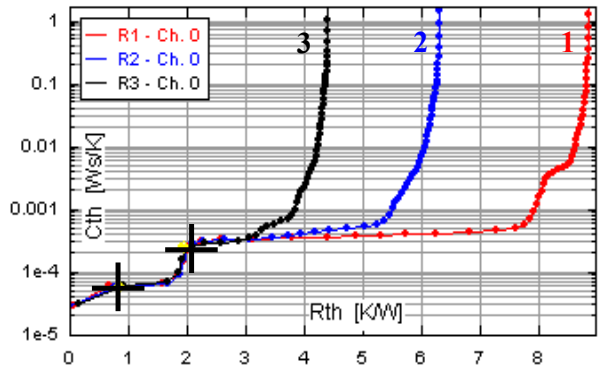

**Figure 16. Structure functions of the three characteristic samples calculated from the thermal transient curves with good sensitivities, two points at the early section for linear fitting**

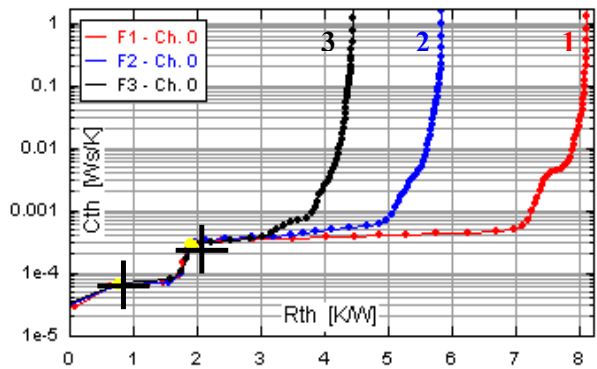

**Figure 17. Structure functions of the three characteristic samples fitted to the two points at the early section**

## 6. CONCLUSIONS

In this paper we presented our investigations on how much time is needed for the thermal transient measurements to achieve short time die attach characterization to enable in-line die attach testing. We examined the question with simulations, calculations and measurements. It was demonstrated that decision about the die attach quality can be made within some 10ms.

To achieve such short measurements we have to eliminate the need of the individual calibration of the temperature sensitive element, in most cases the diode. We showed how the calibration of the temperature dependent parameter of each device can be replaced by applying a linear fitting method in the early range of the measured results.

Using this method the die attach characterisation may become extremely fast. With these investigations it is proven that the structure function based die attach qualification method is applicable for measuring die attach differences between samples from the same type in the fabrication process in some 10ms time interval.

## 7. REFERENCES

[1] M. Rencz, V. Székely, B .Courtois, L. Zhang, N. Howard, L. Nguyen: "Die attach quality control of 3D stacked dies", Proceeding of the IEMT Symposium of SEMICON West, San Jose, Jul 12-16, CA, USA, 2004, Proceedings pp 78-84.

[2] Li Zhang, Noella Howard, Vijaylaxmi Gumaste, Amindya Poddar, Luu Nguyen: "Thermal Characterization of Stacked-Die Packages", 20th IEEE SEMI-THERM Symposium, pp. 55-63

[3] G. Farkas, Q. van Voorst Vader, A. Poppe, Gy. Bognár "Thermal investigation of high power optical devices by transient testing", Proc. of the 9th THERMINIC, Aix-en-Provence, France, Sep 2003

[4] V. Szekely and Tran Van Bien: "Fine structure of heat flow path in semiconductor devices: a measurement and identification method", Solid-State Electronics, V.31, pp. 1363-1368 (1988)

[5] Péter Szabó, Márta Rencz, Gábor Farkas, András Poppe: "Short time die attach characterization of LEDs for in-line testing application", Electornics Packaging Technologies Conference, Singapore, 6-8 December 2006

[6] V. Székely, A. Páhi, M. Rosenthal, M. Rencz: "SUNRED: a field solver and compact model generator tool based on successive mode reduction", MSM'99, April 19-21, 1999, San Juan, Puerto Rico, USA, pp. 342-345

[7] O. Steffens, P. Szabo, M. Lenz, G. Farkas: "Junction to case characterization methodology for single and multiple chip structures based on thermal transient measurements", Proceedings of the XXIth SEMI-THERM Symposium, San Jose, CA, USA, 14-16 March 2005.